\pdfoutput=1

\documentclass[11pt,twoside,a4paper,cmspaper,final,collab]{cms-tdr}
\def\svnVersion{49480M}\def\cmsCernNo{2011-027}\def\cmsCernDate{\today}\def\cmsMessage{Submitted to Physical Review Letters}

\begin{document}\cmsNoteHeader{HIG-10-002}

\hyphenation{had-ron-i-za-tion}
\hyphenation{cal-or-i-me-ter}
\hyphenation{de-vices}
\RCS$Revision: 48532 $
\RCS$HeadURL: svn+ssh://alverson@svn.cern.ch/reps/tdr2/papers/HIG-10-002/trunk/HIG-10-002.tex $
\RCS$Id: HIG-10-002.tex 48532 2011-04-04 13:56:56Z dasu $
\newcommand{\xsecbr}{\sigma_\phi\cdot B_{\tau\tau}}
\newcommand{\MT}{M_{\rm T}}
\cmsNoteHeader{HIG-10-002} 
\title{Search for Neutral MSSM Higgs Bosons Decaying to Tau Pairs in $\Pp\Pp$ Collisions at $\sqrt{s}=7$ TeV}

\address[neu]{Northeastern University}
\address[fnal]{Fermilab}
\address[cern]{CERN}
\author[cern]{The CMS Collaboration}

\date{\today}

\abstract{
A search for neutral MSSM Higgs bosons in $\Pp\Pp$ collisions at the
LHC at a center-of-mass energy of 7~TeV is presented.  The results are
based on a data sample corresponding to an integrated luminosity of
36~pb$^{-1}$ recorded by the CMS experiment.  The search uses decays
of the Higgs bosons to tau pairs.  No excess is observed in the
tau-pair invariant-mass spectrum.  The resulting upper limits on the
Higgs boson production cross section times branching fraction to tau
pairs, as a function of the pseudoscalar Higgs boson mass, yield
stringent new bounds in the MSSM parameter space.}

\hypersetup{%
pdfauthor={CMS Collaboration},%
pdftitle={Search for Neutral MSSM Higgs Bosons Decaying to Tau Pairs in pp Collisions at sqrt(s)=7 TeV},%
pdfsubject={CMS},%
pdfkeywords={CMS, Higgs}}

\maketitle 

The standard model (SM) has been extremely successful in describing a
wide range of phenomena in particle physics, and has survived some
four decades of experimental testing.  However, the only remaining
undiscovered particle predicted by the SM, the Higgs
boson~\cite{Higgs1964-1,Higgs1964-2,Englert,Guralnik,Higgs1966},
suffers from quadratically divergent self-energy corrections at high
energies~\cite{Witten-hierarchy}.  Numerous extensions to the SM have
been proposed to address these divergences.  One such model,
supersymmetry~\cite{Martin97}, a symmetry between fundamental bosons
and fermions, results in cancellation of the divergences at tree
level.  The minimal supersymmetric extension to the standard model
(MSSM) requires the presence of two Higgs doublets.  This leads to a
more complicated Higgs boson sector, with five massive Higgs bosons: a
light neutral scalar ($h$), two charged scalars ($H^\pm$), a heavy
neutral CP-even state ($H$) and a neutral CP-odd state ($A$).

The masses of the MSSM Higgs boson states are specified up to
radiative corrections mainly by two parameters, usually taken to be
the mass of the pseudoscalar state, $m_A$, and the ratio of the vacuum
expectation values of the two Higgs doublets, $\tan\beta$.  At large
$\tan\beta$ (greater than about 20--30), the couplings of the Higgs
bosons to down-type quarks are proportional to $\tan\beta$.  As a
result, the production cross section for two of the three neutral
Higgs bosons can be nearly as large as that for the electroweak gauge
bosons $\PW$ and $\cPZ$ at a proton-proton collider such as the Large
Hadron Collider (LHC).  Two main production processes contribute to
$\Pp\Pp\to\phi+X$, where $\phi=h,H$, or $A$: gluon fusion through a
$\cPqb$ quark loop and direct $\cPqb\cPaqb$ annihilation from the
$\cPqb$ parton density in the beam protons.

The mass relations among the neutral MSSM Higgs bosons are such that
if $m_A \lesssim 130$ GeV/$c^2$, at large $\tan\beta$ the masses of
the $h$ and $A$ are nearly degenerate, while that of the $H$ is
approximately 130 GeV/$c^2$.  If $m_A \gtrsim 130$ GeV/$c^2$, then the
masses of the $A$ and $H$ are nearly degenerate, while that of the $h$
remains near 130 GeV/$c^2$. The precise value of the crossover point
depends predominantly on the nature of the mass mixing in the
top-squark states.

This Letter reports a search for MSSM neutral Higgs bosons in $\Pp\Pp$
collisions at $\sqrt{s}=7$~TeV at the LHC, using a data sample
collected in 2010 corresponding to 36 pb$^{-1}$ of integrated
luminosity recorded by the Compact Muon Solenoid (CMS) experiment.
This search is similar to those performed at the
Tevatron~\cite{Tevatron-MSSM} and complementary to the MSSM Higgs
search at LEP~\cite{LEP2-MSSM}.

The tau-pair decays of the neutral Higgs bosons, having a branching
fraction of about 10\%, serve as the best experimental signature for
this search.  The $\cPqb\cPaqb$ mode, though it has a much larger
branching fraction, suffers from an overwhelming background from QCD
processes.  Three final states where one or both taus decay
leptonically are used: $\Pe\Pgt_h$, $\Pgm\Pgt_h$, and $\Pe\Pgm$, where
we use the symbol $\Pgt_h$ to indicate a reconstructed hadronic decay
of a $\Pgt$.

The central feature of the CMS apparatus is a superconducting
solenoid, of 6~m internal diameter, providing a field of 3.8~T. Within
the field volume are the silicon pixel and strip tracker, the crystal
electromagnetic calorimeter and the brass/scintillator hadron
calorimeter. Muons are measured in gas-ionization detectors embedded
in the steel return yoke. In addition to the barrel and endcap
detectors, CMS has extensive forward calorimetry. Details of the CMS
detector and its performance can be found elsewhere~\cite{CMS-JINST}.

CMS uses a right-handed coordinate system, with the origin at the
nominal interaction point, the $x$ axis pointing to the center of the
LHC, the $y$ axis pointing up (perpendicular to the LHC plane), and
the $z$ axis along the anticlockwise-beam direction. The polar angle
$\theta$ is measured from the positive $z$ axis and the azimuthal
angle $\phi$ is measured in the $xy$ plane.  We measure the
pseudorapidity $\eta$ of outgoing particles based on their polar angle
according to $\eta \equiv -\ln(\tan\frac{\theta}{2}$).

The triggers used to select the events for this analysis are based on
the presence of an electron and/or a muon trigger
object~\cite{CMS-PAS-EGM-10-004,CMS-PAS-MUO-10-002}.  With increasing
instantaneous luminosity, in order to keep the online transverse
momentum thresholds on electrons lower than those used in offline
selections, special triggers requiring the presence of both a lepton
and a charged track with an accompanying calorimeter pattern
consistent with a $\Pgt$ decaying hadronically were adopted for the
$\Pe\Pgt_h$ and $\Pgm\Pgt_h$ channels.

The analysis presented here makes use of particle flow techniques
which combine the information from all CMS subdetectors to identify
and reconstruct individual particles in the event, namely muons,
electrons, photons, and charged and neutral hadrons. The detailed
description of the algorithm and its commissioning can be found
elsewhere~\cite{CMS-PAS-PFT-09-001,CMS-PAS-PFT-10-002}. The particle
list is given as input to the jet, tau, and missing transverse energy
reconstruction.

The main challenge in the identification of hadronic tau decays is
overcoming the large background due to hadronic jets from QCD
processes.  Hadronic tau decays almost always yield one or three
charged pions, plus zero to several neutral pions, depending on the
decay mode.  The algorithm used here starts with a
high-transverse-momentum ($\pt$) reconstructed charged hadron, and
combines it with other nearby reconstructed charged hadron and neutral
pion candidates.  The algorithm considers all possible combinations of
these objects and determines which are consistent with the kinematics
of tau decay.  Among those, it chooses the most isolated in terms of
the presence of nearby reconstructed particles.  Requirements on the
isolation variables determine an operating point in the space of tau
identification efficiency versus the jet-to-tau misidentification
rate.  We optimize the full analysis for best sensitivity by choosing
the ``loose'' operating point of the HPS algorithm~\cite{CMS-PAS-TAU-11-001}.

For the $\Pgm\Pgt_h$ and $\Pe\Pgt_h$ final states, we select events with
an isolated muon or electron with $\pt >$ 15 GeV/$c$ and $|\eta| <
2.1$, and an oppositely charged $\Pgt_h$ with $\pt > 20\GeV /c$ and
$|\eta| < 2.3$.  The transverse mass of the $\ell=\Pe,\Pgm$ with the
missing transverse energy $\ETslash$\ \ , obtained using all
reconstructed particles in the event, is defined as $\MT = \sqrt{ 2
\pt^{\ell} \ETslash \cdot \left ( 1 - \cos {\Delta \phi} \right
) } $, where $\Delta\phi$ is the difference in azimuth between the $\Pe$
or $\Pgm$ and the $\ETslash$ \ vector.  We require $\MT<40$
GeV/$c^2$, in order to reduce the background from $W$+jets events.
For the $\Pe\Pgm$ final state, we select events with an isolated electron
with $|\eta|<2.5$ and an oppositely charged isolated muon, both with
$\pt >$ 15 GeV/$c$ and $\MT <$ 50 GeV/$c^2$, calculated for each
lepton separately.  We reject events in which there are more than one
$\Pe$ or $\Pgm$.

The observed number of events in each channel appears in
Table~\ref{tab-bkg}.  The largest source of events selected with these
requirements comes from $\cPZ\to\Pgt\Pgt$.  We estimate the
contribution from this process using a detailed {\sc geant4}
simulation of the CMS detector, with the events modeled by the {\sc
powheg} Monte Carlo generator~\cite{POWHEG-V,POWHEG1,POWHEG2,POWHEG3}.
We determine the normalization for this process based on the number of
observed $\cPZ\to\ell\ell$ events~\cite{CMS-EWK-WZ}.

A significant source of background arises from QCD multijet events and
$\PW$+jets events in which a jet is misidentified as $\Pgt_h$, and
there is a real or misidentified $\Pe$ or $\Pgm$. The rates for these
processes are estimated using the number of observed same-charge
events, and cross-checked using the jet-to-tau misidentification rate
measured in multijet events.  Other background processes include
$\cPqt\cPaqt$ production and $\cPZ\to \Pe\Pe/\Pgm\Pgm$ events,
particularly in the $\Pe\Pgt_h$ channel, due to the 2--3\% probability
for electrons to be misidentified as
$\Pgt_h$~\cite{CMS-PAS-TAU-11-001}.  The small fake-lepton background
from $\PW +$jets and QCD for the $\Pe\Pgm$ channel is estimated using
data.  Table~\ref{tab-bkg} shows the expected number of events for
each of the background processes.  The event generator {\sc
pythia6}~\cite{PYTHIA6} is used to model the Higgs boson signal and
other backgrounds.  The {\sc tauola}~\cite{TAUOLA} package is used for
tau decays in all cases.

\begin{table}
  \begin{center}
    \caption{Number of estimated background events in the selected
             sample, observed number of events, and overall signal efficiency for $m_A=200$ GeV/$c^2$
             (including all branching fractions) for each search channel.  Uncertainties do not
             include those on integrated luminosity or energy scales.  The QCD
             multijet background for the $\Pe\Pgt_h$ final state is the
             sum of the QCD multijet and $\gamma$+jet backgrounds.}
    \begin{tabular}{lccc}\hline\hline
Process                                                       & $\Pgm\Pgt_h$     & $\Pe\Pgt_h$      & $\Pe\Pgm$      \\ \hline \hline
$\cPZ \to \Pgt\Pgt$                                           & $  329\pm 77 $   & $  190\pm 44 $   & $  88\pm 5  $  \\ \hline
$\cPqt\cPaqt$                                                 & $    6\pm 3  $   & $  2.6\pm 1.3$   & $ 7.1\pm 1.3$  \\ \hline
$\cPZ \rightarrow \ell \ell$, ${\rm jet} \rightarrow \Pgt_h$  & $  6.4\pm 2.4$   & $   15\pm 6.2$   &      -         \\ \hline
$\cPZ \rightarrow \ell \ell$                                  & $ 12.9\pm 3.5$   & $  109\pm 28 $   & $ 2.4\pm 0.3$  \\ \hline
$\PW \rightarrow \ell \cPgn$                                  & $ 54.9\pm 4.8$   & $ 30.6\pm 3.1$   &                \\ \cline{1-3}
$\PW \rightarrow \Pgt\cPgn, \Pgt\to\ell\cPgn\bar{\cPgn}$      & $ 14.7\pm 1.3$   & $  7.0\pm 0.7$   & $ 1.5\pm 0.5$  \\ \cline{1-3}
QCD multijet and $\Pgg$+jet                                   & $  132\pm 14 $   & $  181\pm 23 $   &                \\ \hline
$\PW\PW/\PW\cPZ/\cPZ\cPZ$                                     & $  1.6\pm 0.8$   & $  0.8\pm 0.4$   & $ 3.0\pm 0.4$  \\ \hline\hline
Total                                                         & $  557\pm 79 $   & $  536\pm 57 $   & $ 102\pm 5 $   \\ \hline
Observed                                                      &        517       &     540          &    101         \\ \hline\hline
Signal Efficiency                                             &    0.0391        &   0.0245         &   0.00582      \\ \hline\hline
    \end{tabular}
    \label{tab-bkg}
  \end{center}
\end{table}

\begin{figure}[hbtp]
  \begin{center}
   \includegraphics[width=0.5\textwidth]{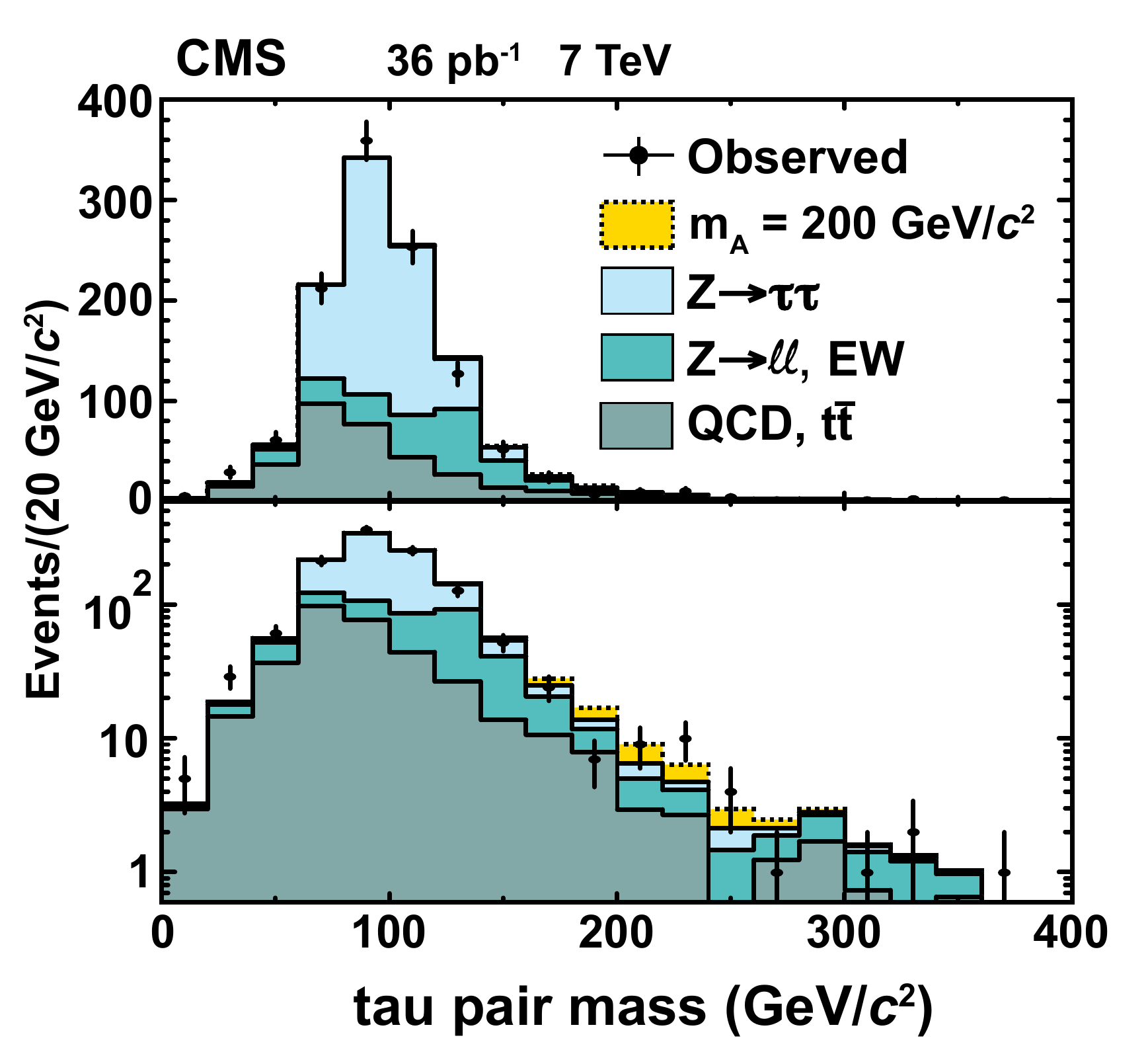}
    \caption{The reconstructed tau-pair invariant-mass distribution on linear (above) and
             logarithmic (below) scales, for the sum of the
             $\Pe\Pgt_h$, $\Pgm\Pgt_h$, and $\Pe\Pgm$ final states, comparing the
             observed distributions (points with error bars) to the sum
             of the expected backgrounds (shaded histograms).  The
             contribution from a Higgs boson signal ($m_A=200$ GeV/$c^2$)
             is also shown, with normalization corresponding to the 95\% upper
             bound on $\xsecbr$.}
    \label{tau-pair-mass}
  \end{center}
\end{figure}

To distinguish the Higgs boson signal from the background, we
reconstruct the tau-pair mass using a likelihood technique.  The
algorithm estimates the original tau three-momenta by maximizing a
likelihood with respect to free parameters corresponding to the
missing tau-neutrino momenta, and subject to all applicable kinematic
constraints.  Other terms in the likelihood take into account the
tau-decay phase space and the probability density in the tau
transverse momentum, parametrized as a function of the tau-pair mass.
This algorithm yields a tau-pair mass with a mean consistent with
the true value, and a distribution with a nearly Gaussian shape.  The
mass resolution is $\sim$21\% at a Higgs boson mass of 130 GeV/$c^2$, to be
compared with $\sim$24\% for the (non-Gaussian) distribution of the
invariant mass reconstructed from the visible tau-decay products.  The
observed reconstructed tau-pair mass distribution for all three
channels is shown in Fig.~\ref{tau-pair-mass}.

Various imperfectly known or imperfectly simulated effects can alter
the shape and normalization of the reconstructed tau-pair
invariant-mass spectrum.  The main sources of normalization
uncertainties include the total integrated luminosity
(11\%)~\cite{CMS-LUMI}, background normalizations
(Table~\ref{tab-bkg}), $\cPZ$ production cross section (4\%), and lepton
identification and isolation efficiency (0.2--2.0\% depending on
lepton type).  The tau identification efficiency uncertainty is
estimated to be 23\% from an independent study~\cite{CMS-PAS-TAU-11-001}.
Uncertainties that contribute to mass spectrum shape variations
include the tau (3\%), muon (1\%), and electron (2\%) energy scales,
and uncertainties on the $\ETslash$ \ scale that is used for the
tau-pair invariant-mass reconstruction~\cite{CMS-EWK-TAU}. The
$\ETslash$ \ scale uncertainties contribute via the jet-energy
scale (3\%) and unclustered energy scale (10\%), where the unclustered
energy is defined as the energy remaining after vectorially
subtracting leptons and objects clustered in jets with $\pt >10 \GeVc$.

To search for the presence of a Higgs boson signal in the selected
events, we perform a maximum likelihood fit to the tau-pair
invariant-mass spectrum.  Systematic uncertainties are represented by
nuisance parameters, which we remove by marginalization, assuming a
log normal prior for normalization parameters, and Gaussian priors for
mass-spectrum shape uncertainties.  The uncertainties that affect the
shape of the mass spectrum, mainly those corresponding to the energy
scales, are represented by nuisance parameters whose variation results
in a continuous modification of the spectrum
shape~\cite{Conway-PhyStat}.

The parameter representing the tau identification uncertainty affects
taus from the Higgs boson signal and the main background,
$Z\to\Pgt\Pgt$, equally.  This effectively allows the observed
$Z\to\Pgt\Pgt$ events to provide an {\em in situ} calibration of this
efficiency, except for Higgs boson masses near that of the $\cPZ$.  Near the
$\cPZ$ mass, the tau identification efficiency uncertainty dominates in
the $\Pe\Pgt_h$ and $\Pgm\Pgt_h$ channels, and the $\Pe\Pgm$ channel thus
provides the greatest sensitivity.

The mass spectra show no evidence for the presence of a Higgs boson
signal, and we set 95\% CL (confidence level) upper bounds on the
Higgs boson cross section times the tau-pair branching fraction
(denoted by $\xsecbr$) using a Bayesian method assuming a uniform
prior in $\xsecbr$.  The invariant-mass spectrum in
Fig.~\ref{tau-pair-mass} shows the result of a fit with a Higgs boson
signal corresponding to $m_A$ = 200 GeV/$c^2$ present, for
$\xsecbr=8.71$ pb, the value above which we exclude at 95\% CL.

\begin{figure}[hbtp]
  \begin{center}
    \includegraphics[width=0.5\textwidth]{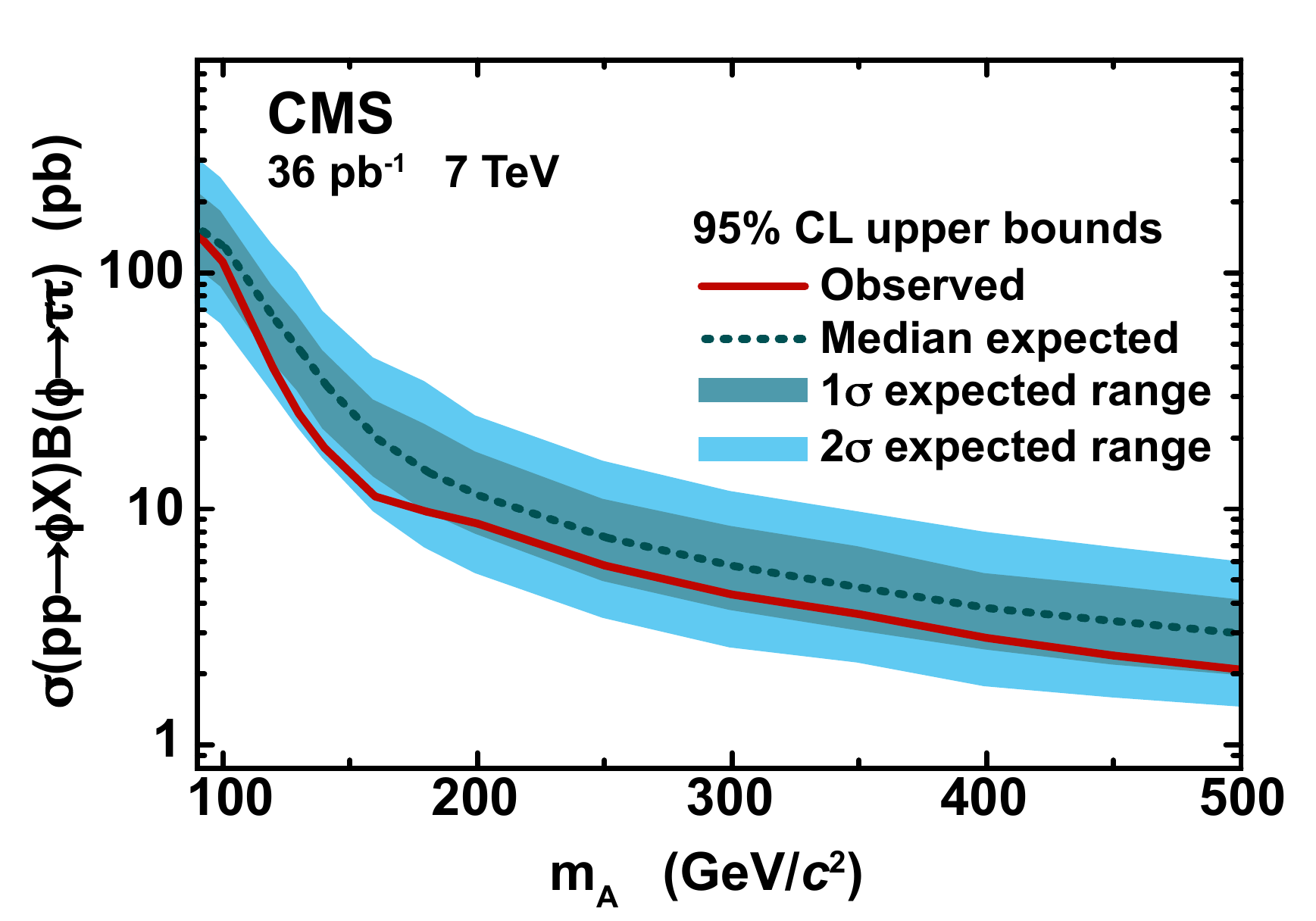}\hspace{1cm}
    \caption{The expected one- and two-standard-deviation ranges and observed 95\%  CL upper limits on
    $\xsecbr$ as a function of $m_A$.}
    \label{xsec-limit}
  \end{center}
\end{figure}

Figure~\ref{xsec-limit} shows the observed upper bound on $\xsecbr$ as
a function of $m_A$, where we use as the signal acceptance model the
combined mass spectra from the $\Glu\Glu$ and $\cPqb\cPaqb$
production processes for $h$, $A$, and $H$, and assuming
$\tan\beta=30$~\cite{LHC-HCWG}.  The plot also shows the one- and
two-standard-deviation range of expected upper limits for various
potential experimental outcomes.  The observed limits are well within
the expected range assuming no signal.  The observed and expected
upper limits are shown in Table~\ref{tab-limits}.

\begin{table}
  \begin{center}
    \caption{Expected range and observed 95\% CL upper limits for
             $\xsecbr$ as functions of $m_A$, and 95\% CL upper
             bound on $\tan\beta$ in the $m_h^{\rm max}$
             scenario described in the text.  No bounds on $\tan\beta$
             above 60 are quoted.}
    \begin{tabular}{cccccccc} \hline \hline
              &    \multicolumn{5}{c}{ 95\% CL Upper Limit}   \\ \hline
   $m_A$      &    \multicolumn{3}{c}{Expected $\xsecbr$ (pb)}  & \multicolumn{2}{c}{Observed} \\
 (GeV/$c^2$)     &$-1\sigma$& Median   &$+1\sigma$&$\xsecbr$ & $\tan\beta$ \\ \hline
  90  &     107.75  &  153.30  &  227.10  &  147.74  &   27.36  \\
 100  &      88.61  &  127.09  &  184.17  &  112.30  &   29.20  \\
 120  &      42.72  &   62.48  &   90.24  &   39.61  &   25.22  \\
 130  &      31.97  &   45.96  &   67.11  &   25.40  &   22.57  \\
 140  &      22.14  &   32.81  &   47.30  &   18.20  &   23.55  \\
 160  &      13.83  &   19.70  &   29.27  &   11.37  &   23.83  \\
 180  &       9.95  &   14.16  &   23.13  &    9.78  &   28.08  \\
 200  &       7.90  &   11.36  &   17.61  &    8.71  &   32.98  \\
 250  &       5.01  &    7.54  &   11.15  &    5.77  &   43.35  \\
 300  &       3.77  &    5.71  &    8.58  &    4.36  &   56.62  \\
 350  &       3.09  &    4.64  &    7.04  &    3.60  &   -      \\
 400  &       2.57  &    3.79  &    5.39  &    2.86  &   -      \\
 450  &       2.21  &    3.34  &    4.77  &    2.41  &   -      \\
 500  &       2.00  &    2.95  &    4.18  &    2.10  &   -      \\ \hline \hline
    \end{tabular}
    \label{tab-limits}
  \end{center}
\end{table}

\begin{figure}[hbtp]
  \begin{center}
    \includegraphics[width=0.5\textwidth]{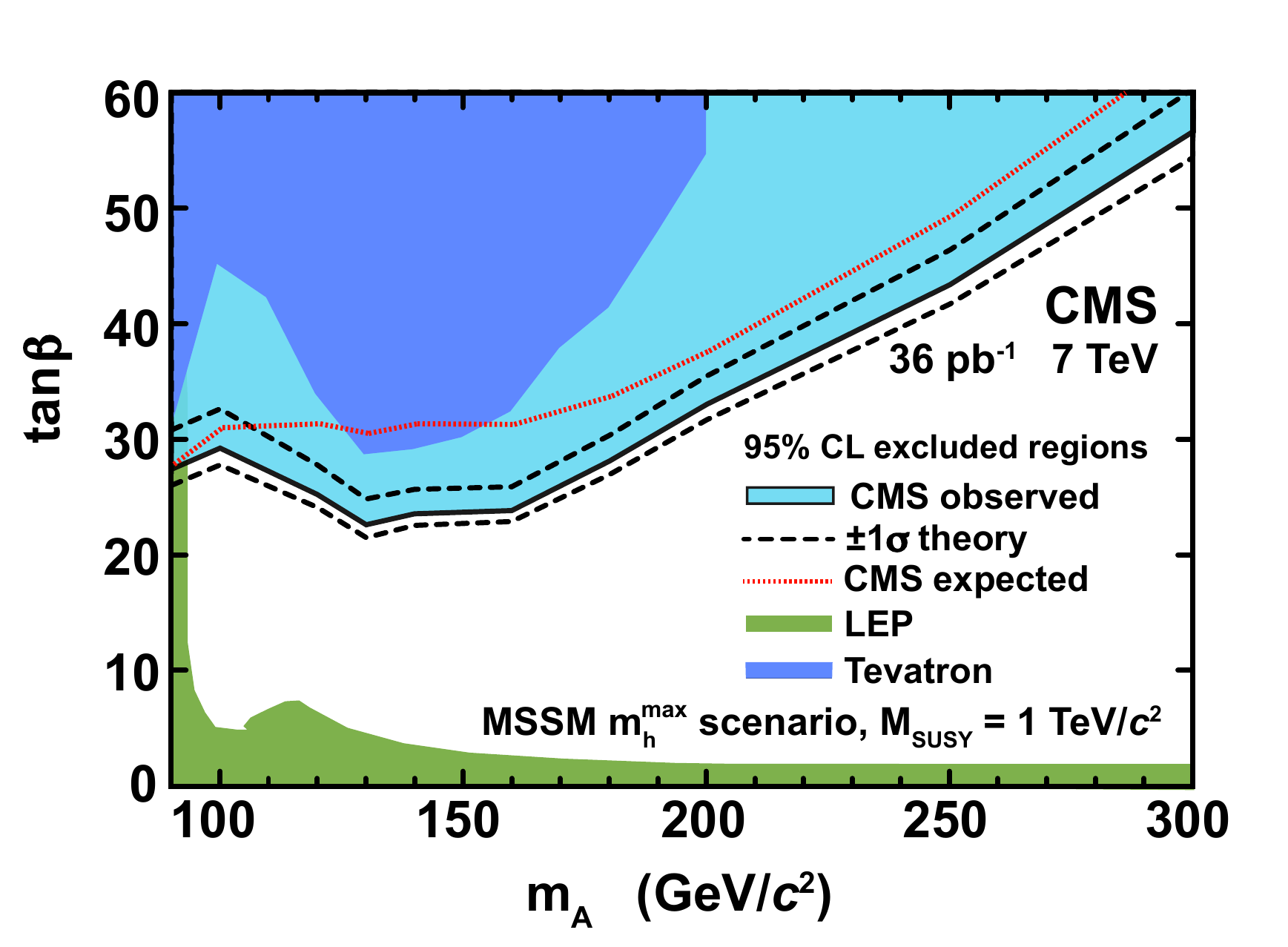} \hspace{1cm}
    \caption{Region in the parameter space of $\tan\beta$ versus $m_A$ excluded
             at 95\% CL in the context of the MSSM $m^{\rm max}_h$ scenario, with
             the effect of $\pm 1\sigma$ theoretical uncertainties shown.  The
             other shaded regions show the 95\% CL excluded regions from the
             LEP and Tevatron experiments.}
    \label{tanbeta-ma}
  \end{center}
\end{figure}

We can interpret the upper limits on $\xsecbr$ in the MSSM parameter
space of $\tan\beta$ versus $m_A$ for an example scenario.  We use
here the $m_h^{\rm max}$~\cite{MHMAX-Carena} benchmark scenario in
which $M_{\rm SUSY}$ = 1~TeV/$c^2$; $X_t$ = 2$M_{\rm SUSY}$; $\Pgm$ =
200 GeV/$c^2$; $M_{\tilde{g}}$ = 800 GeV/$c^2$; $M_2$ = 200 GeV/$c^2$;
and $A_b = A_t$, where $M_{\rm SUSY}$ denotes the common
soft-SUSY-breaking squark mass of the third generation; $X_t = A_t
- \Pgm/\tan\beta$ the stop mixing parameter; $A_t$ and $A_b$ the stop
and sbottom trilinear couplings, respectively; $\Pgm$ the Higgsino mass
parameter; $M_{\tilde{g}}$ the gluino mass; and $M_2$ the
SU(2)-gaugino mass parameter. The value of $M_1$ is fixed via the GUT
relation $M_1 =(5/3)M_2\sin\theta_{\rm W}/\cos\theta_{\rm W}$.  In
determining these bounds on $\tan\beta$, shown in
Table~\ref{tab-limits} and in Fig.~\ref{tanbeta-ma}, we have used the
central values of the Higgs boson cross sections as a function of
$\tan\beta$ reported by the LHC Higgs Cross Section Working
Group~\cite{LHC-HCWG}.  The cross sections have been obtained from the
{\sc GGH@NNLO}~\cite{Harlander:2002wh,Harlander:2002vv} and {\sc
HIGLU}~\cite{Spira:1995mt} programs for the gluon-fusion process and
from the {\sc BBH@NNLO}~\cite{Harlander:2003ai} program for the
$\cPqb\cPaqb\to \phi$ process in the 5-flavor scheme, rescaling the
corresponding Yukawa couplings by the MSSM factors calculated with
FeynHiggs~\cite{Heinemeyer:1998yj}.  The $\Glu\Glu\to\phi$ cross-section
calculations combine the full quark mass-dependent NLO QCD
corrections~\cite{Spira:1995rr} and NNLO corrections in the
heavy-top-quark
limit~\cite{Harlander:2002wh,Anastasiou:2002yz,Ravindran:2003um}.  The
effect of the theoretical uncertainties is illustrated in
Fig.~\ref{tanbeta-ma}.  We do not quote limits above $\tan\beta = 60$
as the theoretical relation between cross section and $\tan\beta$
becomes unreliable.

The present results exclude a region in $\tan\beta$ down to values
smaller than those excluded by the Tevatron
experiments~\cite{Tevatron-MSSM} for $m_A\lesssim$ 140 GeV/$c^2$, and
significantly extend the excluded region of MSSM parameter space at
larger values of $m_A$.  Figure~\ref{tanbeta-ma} also shows the region
excluded by the LEP experiments~\cite{LEP2-MSSM}.

In conclusion, we have performed a search for neutral MSSM Higgs
bosons, using the first sample of CMS data from proton-proton
collisions at a center-of-mass energy of 7~TeV at the LHC,
corresponding to an integrated luminosity of 36~pb$^{-1}$.  The
tau-pair decay mode in final states with one $\Pe$ or $\Pgm$ plus a
hadronic decay of a tau, and the $\Pe\Pgm$ final state were used.  The
observed tau-pair mass spectrum reveals no evidence for neutral Higgs
boson production, and we determine an upper bound on the product of
the Higgs boson cross section and tau-pair branching fraction as a
function of $m_A$.  These results, interpreted in the MSSM parameter
space of $\tan\beta$ versus $m_A$, in the $m_h^{\rm max}$ scenario,
exclude a previously unexplored region reaching as low as $\tan\beta =
23$ at $m_A = 130$ GeV/$c^2$.

We wish to congratulate our colleagues in the CERN accelerator
departments for the excellent performance of the LHC machine. We thank
the technical and administrative staff at CERN and other CMS
institutes, and acknowledge support from: FMSR (Austria); FNRS and FWO
(Belgium); CNPq, CAPES, FAPERJ, and FAPESP (Brazil); MES (Bulgaria);
CERN; CAS, MoST, and NSFC (China); COLCIENCIAS (Colombia); MSES
(Croatia); RPF (Cyprus); Academy of Sciences and NICPB (Estonia);
Academy of Finland, ME, and HIP (Finland); CEA and CNRS/IN2P3
(France); BMBF, DFG, and HGF (Germany); GSRT (Greece); OTKA and NKTH
(Hungary); DAE and DST (India); IPM (Iran); SFI (Ireland); INFN
(Italy); NRF and WCU (Korea); LAS (Lithuania); CINVESTAV, CONACYT,
SEP, and UASLP-FAI (Mexico); PAEC (Pakistan); SCSR (Poland); FCT
(Portugal); JINR (Armenia, Belarus, Georgia, Ukraine, Uzbekistan); MST
and MAE (Russia); MSTD (Serbia); MICINN and CPAN (Spain); Swiss
Funding Agencies (Switzerland); NSC (Taipei); TUBITAK and TAEK
(Turkey); STFC (United Kingdom); DOE and NSF (USA).

 \bibliography{auto_generated}   

\cleardoublepage\appendix\section{The CMS Collaboration \label{app:collab}}\begin{sloppypar}\hyphenpenalty=5000\widowpenalty=500\clubpenalty=5000\textbf{Yerevan Physics Institute,  Yerevan,  Armenia}\\*[0pt]
S.~Chatrchyan, V.~Khachatryan, A.M.~Sirunyan, A.~Tumasyan
\vskip\cmsinstskip
\textbf{Institut f\"{u}r Hochenergiephysik der OeAW,  Wien,  Austria}\\*[0pt]
W.~Adam, T.~Bergauer, M.~Dragicevic, J.~Er\"{o}, C.~Fabjan, M.~Friedl, R.~Fr\"{u}hwirth, V.M.~Ghete, J.~Hammer\cmsAuthorMark{1}, S.~H\"{a}nsel, M.~Hoch, N.~H\"{o}rmann, J.~Hrubec, M.~Jeitler, G.~Kasieczka, W.~Kiesenhofer, M.~Krammer, D.~Liko, I.~Mikulec, M.~Pernicka, H.~Rohringer, R.~Sch\"{o}fbeck, J.~Strauss, F.~Teischinger, P.~Wagner, W.~Waltenberger, G.~Walzel, E.~Widl, C.-E.~Wulz
% ...
\vskip\cmsinstskip
% ...
1:~~Also at CERN, European Organization for Nuclear Research, Geneva, Switzerland\\
% ...

\end{sloppypar}
\end{document}